\documentstyle[12pt]{article}
\input{epsf}
\newcommand{\cket}[1]{| #1 \rangle}
\newcommand{\bra}[1]{\langle #1 |}

\begin{document}

\begin{center}
{\large Negative-parity Nucleon Resonance in the QCD Sum Rule \\}
\vspace*{0.5cm}
{ D. Jido , N. Kodama and M. Oka \\
Department of Physics, Tokyo Institute of Technology\\ 
Meguro, Tokyo 152  Japan} 
\end{center} 
\abstract{The negative-parity baryons are studied by a novel approach 
in the QCD sum rule.  It is found that the parity of the 
ground-state nucleon is determined by the sign of the quark condensate. 
We predict the mass of negative-parity nucleon.  }
\section{Introduction}
The QCD sum rule, proposed by Shifman, Vainshtein and 
Zakharov~\cite{svz}, connects hadron properties and QCD 
parameters~\cite{rry}.  
The correlation function of an interpolating field for the 
a hadron is expressed in two ways: (1) OPE side: the 
correlation function is calculated perturbatively at deep Euclidean 
momentum with the help of the operator product expansion, and (2) 
phenomenological side: it is expressed in terms of hadron spectral 
function in the physical region.  Using the analyticity of the 
correlation function the two expressions are connected in the integral 
form, which is the sum rule.  The time-ordered correlation function is 
usually employed, but in this paper from a technical reason we use the 
``old-fashioned'' correlation function defined by
\begin{equation}
   \Pi(p) = i \int d^{4}x e^{ip\cdot x} \theta(x_{0}) \bra{0} J(x) 
   \bar{J}(0) \cket{0}, \label{eq:ourco}
\end{equation}
where $J(x)$ is a local operator which annihilates a hadron and is called
an interpolation current.

The QCD sum rule is applied to baryons by Ioffe~\cite{i}~\cite{i2}.  For the 
octet baryons two independent currents which contain no derivative are 
available.  The general expression of the nucleon current 
is~\cite{ept}
\begin{equation}
   J_{N}^{\alpha}(x) = \varepsilon_{abc} [(u_{a}(x)Cd_{b}(x)) 
      (\gamma_{5} u_{c}(x))^{\alpha} + 
  t (u_{a}(x) C \gamma_{5} d_{b}(x)) u_{c}^{\alpha}(x)], \label{eq:nucur}
\end{equation}
where $u(x)$ and $d(x)$ are field operators of up and down quarks, $C$ 
is the charge conjugation operator and $abc$ are color indices.  
$J_{N}^{\alpha}$ $(\alpha = 1,2,3,4)$ forms a Dirac spinor.  A commonly used 
current assumes $t = -1$ in eq.(~\ref{eq:nucur}), which is called 
``Ioffe's current''.  It is optimal for the lowest-lying nucleon\cite{ept},
i.e.\ it couples strongly to the lowest-lying nucleon state.
Because, as we shall see later, the current for baryons couples also 
to the negative-parity baryons~\cite{cdks}, other choices of $t$ enable us
to study the negative-parity baryons.  In this paper, we extract the mass 
of the negative-parity nucleon from the sum rule with the nucleon current
by choosing $t$ so that the current strongly couples to the negative-parity 
nucleon.  A similar study has been done by the authors of ref.~\cite{cdks}.
they, however, used the current that contains an operator with derivative.
They did not find a Boral stability in the prediction of
the negative-parity nucleon mass and therefore the results are only
qualitative.

In sect.2, we study the relation between the positive and negative 
parity baryons in the sum rule.  We point out that the sum rule for 
the positive-parity baryon contains the contribution of the 
negative-parity baryons.  We propose a technique to separate the 
contribution of the negative-parity baryons from the sum rule in 
sect.3.  In sect.4, we apply our formulation to the negative-parity 
nucleon resonance, and calculate its mass.  We study the $t$ 
dependence (eq.(~\ref{eq:nucur})) of the sum rule in detail.  We find 
that the sign of the quark condensate determines the order of the 
parity doublet.  In sect.5, roles of the chiral symmetry breaking 
quark condensates are studied in the masses of the positive- and 
negative-parity nucleons.  We study the behavior of the masses when the 
quark condensates are varied.  A summary is given in sect.6. 

\section{Negative-parity Baryons in the QCD Sum Rule}
A baryon current studied in the QCD sum rule is composed of three 
quark fields, and it couples to the states with the same quantum 
number as the current.  In the mesonic case, the parity of the meson 
that couples to the current is directly connected to the parity of the 
current.  That is, the parity of the meson coincides with the parity 
of the bilinear form, $\bar{q} \Gamma q$.  For instance, the current 
for the $\rho^{+}$-meson is $\bar{d}\gamma_{\mu} u$, while the current 
for $a_{1}^{+}$-meson, which is the chiral partner of $\rho$, is 
$\bar{d}\gamma_{\mu}\gamma_{5}u$.  It may seem that the QCD sum rule 
for a negative-parity baryon is similarly given by the current $J_{-} 
\equiv i\gamma_{5} J_{+}$ as an interpolating field because 
multiplying $i \gamma_{5}$ to $J_{+}$ changes the ``parity'' of $J_{+}$, 
where $J_{+}$ is the current for the corresponding positive-parity 
baryon, such as $J_{N}$ in eq.(~\ref{eq:nucur}).  Note that both 
$J_{+}$ and $J_{-}$ are Dirac (4-component) spinors.

Suppose that the correlation function of $J_{+}$ is given by 
\begin{equation}
    \Pi_{+}(p) = p_{\mu}\gamma^{\mu}\Pi_{1}(p^{2}) + \Pi_{2}(p^{2}),
\end{equation}
then the correlation function of $J_{-}$ can be written as 
\begin{equation}
    \Pi_{-}(p) = -\gamma_{5} \Pi_{+}(p)\gamma_{5} = p_{\mu} 
    \gamma^{\mu}\Pi_{1}(p^{2}) - \Pi_{2}(p^{2}).  
\end{equation}
The difference between $J_{+}$ and $J_{-}$ appears only in the sign in 
front of $\Pi_{2}(p^{2})$.  That is, the same functions 
$\Pi_{1}(p^{2})$ and $\Pi_{2}(p^{2})$ appear in $\Pi_{+}(p)$ and 
$\Pi_{-}(p)$.  Because we construct the sum rules separately for 
$\Pi_{1}(p)$ and $\Pi_{2}(p)$, we do not get any independent sum rule 
from $J_{-}$.

In fact, the information about the negative-parity baryons is already 
in $\Pi_{+}(p)$ since $J_{+}$ couples not only to the positive-parity 
baryons but also to the negative-parity baryons~\cite{cdks}.  It is 
easy to see this from
\begin{equation}
    \bra{0}J_{+}\cket{B^{-}} \bra{B^{-}}\bar{J_{+}}\cket{0} = 
    - \gamma_{5}\bra{0}J_{-}\cket{B^{-}}\bra{B^{-}} 
    \bar{J_{-}}\cket{0} \gamma_{5},
\end{equation}
where $\cket{B^{-}}$ denotes a single baryon state with negative parity.  
$J_{-}$ couples to the positive-parity states in the same way.

To see concretely how the information of the negative-parity baryons 
is included in the correlation function $\Pi^{+}(p)$, we express the 
time-ordered correlation function as a sum of contributions from 
zero-width poles:
\begin{eqnarray}
    \Pi^{T}(p) & \equiv & i \int d^{4}x \, e^{ip\cdot x} \bra{0} T 
    J_{+}(x) \bar{J}_{+}(0) \cket{0} \nonumber \\
    & = & \sum_{n} \left[ \lambda_{n}^{+}{\gamma_{\mu} p^{\mu} + 
    m^{+}_{n} \over p^{2} - (m^{+}_{n})^{2}} + \lambda_{n}^{-}
    {\gamma_{\mu} p^{\mu} - m^{-}_{n} \over p^{2} - (m^{-}_{n})^{2}} \right],
\end{eqnarray}
where $m_{n}^{\pm}$ is the mass of the $n$-th resonance and 
$\lambda_{n}^{\pm}$ is the coupling strength of the current to the 
resonance.  Note that only difference of the positive-parity part and 
the negative-parity part is the sign of the mass terms.  When we are 
interested in the lowest-lying baryon with positive parity, we regard 
excited states as a part of "continuum" contribution.  Then the terms 
from the negative parity baryons cannot be directly seen.  In the next 
section, we propose a formulation for separating the negative-parity 
contribution from the sum rule.

\section{Sum Rule for the Negative-parity Baryon}
\label{sc:sep}
To separate the terms of negative-parity baryons from those of 
positive parity baryons, we use the ``old-fashioned'' correlation 
function (~\ref{eq:ourco}).  In the zero-width resonance 
approximation, we write the imaginary part in the rest frame $\vec{p} 
= 0$ as
\begin{eqnarray}
  {\rm Im} \, \Pi(p_{0}) & = & \sum_{n} \left[
    (\lambda_{n}^{+})^{2} {\gamma_{0} + 1\over 2} \delta(p_{0} - m_{n}^{+}) + 
    (\lambda_{n}^{-})^{2} {\gamma_{0} - 1\over 2} \delta(p_{0} - m_{n}^{-}) \right] 
  \label{eq:imphe} \\
  & \equiv & \gamma_{0} A(p_{0}) + B(p_{0}) \nonumber,
\end{eqnarray}
where $A(p_{0})$ and $B(p_{0})$ are defined by
\begin{eqnarray*}
  A(p_{0}) & = & {1 \over 2} \sum_{n}[ (\lambda_{n}^{+})^{2} 
  \delta(p_{0} - m_{n}^{+})  + (\lambda_{n}^{-})^{2} \delta(p_{0} - 
  m_{n}^{-})], \\
  B(p_{0}) & = & {1 \over 2} \sum_{n}[ (\lambda_{n}^{+})^{2} 
  \delta(p_{0} - m_{n}^{+}) - (\lambda_{n}^{-})^{2} \delta(p_{0} - 
  m_{n}^{-})] .
\end{eqnarray*}
One can see that the contribution $A(p_{0}) + B(p_{0})$ ($A(p_{0}) - 
B(p_{0})$) contains contributions only from the positive-parity 
(negative-parity) states.

We, however, can no longer construct sum rules in $p^{2}$-space, since 
the ``old-fashioned'' correlation function is not analytic in $p^{2}$ 
space.  Instead a sum rule can be written in the complex $p_{0}$ 
plane, because the correlation function (~\ref{eq:ourco}) is analytic 
in the upper-half region of the complex $p_{0}$ plane.  The 
theoretical side is given by the operator product expansion, which is 
valid at high energy i.e.  $\Pi^{\rm OPE}(p_{0}=Q) \simeq \Pi^{\rm 
Phe}(p_{0}=Q)$ at large $|Q|$.  Using the analyticity we obtain 
independent sum rules
\begin{eqnarray}
    \int_{0}^{Q}[A^{\rm OPE}(p_{0}) - A^{\rm Phe}(p_{0}) ] W(p_{0}) \, dp_{0}
        & = & 0, \label{eq:sra} \\
    \int_{0}^{Q}[B^{\rm OPE}(p_{0}) - B^{\rm Phe}(p_{0}) ] W(p_{0}) \, dp_{0} 
        & = & 0,  \label{eq:srb}
\end{eqnarray}
where $W(p_{0})$ is an arbitrary analytic function which is real on 
the real axis.  Note that we use the fact that the imaginary part of the 
correlation vanishes in negative $p_{0}$.

We use the Borel weight $W(p_{0}) = \exp( - \frac{{p_{0}}^{2}}{M^{2}})$.
We take the lowest mass pole and approximate others as 
continuum whose behavior above a threshold $s_{0}^{\pm}$ is same as the 
theoretical side. Then we obtain two sum rules
\begin{eqnarray}
     {1 \over 2} [\tilde{A}^{\rm OPE}(M,s_{0}^{+}) + \tilde{B}^{\rm 
     OPE}(M,s_{0}^{+})] & = & 
     (\lambda^{+})^{2} \exp[{- \frac{(m^{+})^{2}}{M^{2}}}],  \label{eq:srp} \\
     {1 \over 2} [\tilde{A}^{\rm OPE}(M,s_{0}^{-}) - \tilde{B}^{\rm 
     OPE}(M,s_{0}^{-})] & = & 
     (\lambda^{-})^{2} \exp[{- \frac{(m^{-})^{2}}{M^{2}}}], \label{eq:srn}
\end{eqnarray} 
where 
\begin{eqnarray*}
     \tilde{A}^{\rm OPE}(M,s_{0}^{+}) = \int^{s_{0}^{+}}_{0}dp_{0}
     A^{\rm OPE}(p_{0}) \exp(- \frac{p^{2}_{0}}{M^{2}}) , \\
     \tilde{B}^{\rm OPE}(M,s_{0}^{-}) = \int^{s_{0}^{-}}_{0}dp_{0}
     B^{\rm OPE}(p_{0}) \exp(- \frac{p^{2}_{0}}{M^{2}}).
\end{eqnarray*}
The first sum rule is for the baryons with positive parity and the 
second one is for negative-parity baryons.  In these sum rules, we 
allow the threshold to be different for each parity.

There are three remarks.  First, the imaginary part of our correlation 
function is written as $ 1/2 \, ({\rm Im} \, \Pi^{\rm T}(p_{0}) + {\rm 
Im} \, \Pi^{\rm R}(p_{0}))$, where T and R stand for ``time-ordered'' 
and ``retarded'', respectively.  The real parts of time-ordered and 
retarded functions are the same, but the sign of negative energy part 
in their imaginary part is different.  It is due to this difference 
that the time-ordered correlation functions is non-analytic and the 
retarded correlation function is analytic in complex $p_{0}$ plane.  
Second, the retarded correlation function, indeed, is analytic on the 
upper half in complex $p_{0}$ plane.  But we can not construct 
$\Pi^{\rm R}$ sum rule, since the integral of $A(p_{0})$, which is an 
odd function of $p_{0}$, vanishes in the sum rules (~\ref{eq:srp}) and 
(~\ref{eq:srn}).  At last, from (~\ref{eq:srp}) and (~\ref{eq:srn}), 
we see that the term $B$ causes the parity splitting.  $B$ is not 
invariant under the chiral transformation.  We, therefore, confirm 
that the chiral symmetry breaking gives the parity splitting of the 
baryon.

\section{Mass of the Negative-parity Nucleon}
In this section we calculate the mass of the negative parity nucleon 
$N^{-}$, and see that the mass is larger than that of the positive 
parity $N^{+}$.

\begin{figure}[tb]
\epsfxsize=12cm
\epsfbox{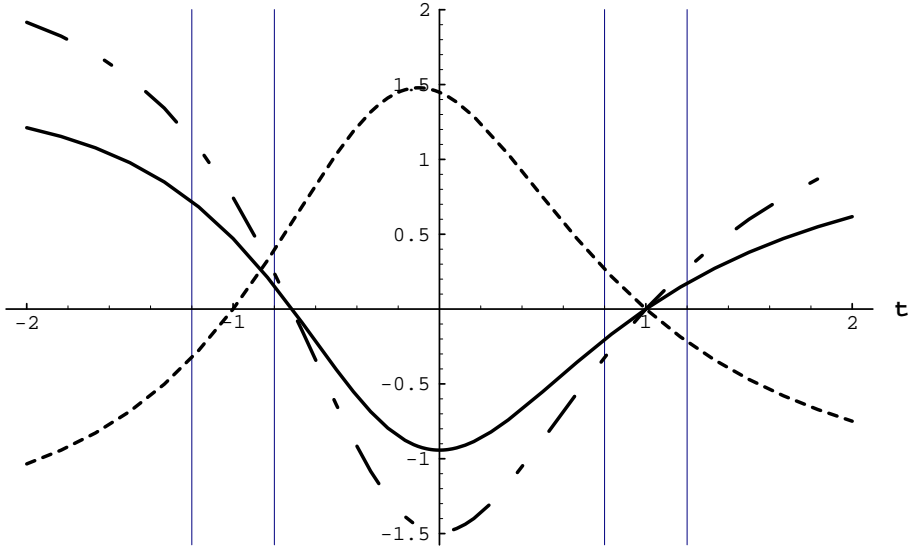}
\caption[]{\small 
   The $t$ dependence of Borel-transformed power corrections divided by 
   the Wilson coefficient of identity operator at the Borel mass 1.5 
   GeV.  The dot-dashed line is the ratio of the Wilson coefficients of 
   $\langle \bar{q} q \rangle$ and identity operator.  The dashed line is 
   for $\langle \bar{q} \sigma \cdot G q \rangle $.  The solid line is 
   the $\langle (\bar{q} q)^{2} \rangle $.}
   \label{fg:t-dep} 
\end{figure}

Using the current (~\ref{eq:nucor}), the theoretical (OPE) side of the 
sum rules (~\ref{eq:sra}) and (~\ref{eq:srb}) up to dimension 6 
operators are given by
\begin{eqnarray}
    {\rm Im} A^{\rm OPE}(p_{0}) & = & {5 + 2 t + 5 t^2 \over 2^{11} 
    \pi^{4}} p_{0}^{\,5} 
    \theta (p_{0})
    + { 5 + 2 t + 5 t^{2} \over 2^{9} \pi^{2}} p_{0} \theta (p_{0}) \langle 
    {\alpha_{s} \over \pi} GG \rangle + \nonumber \\
    & & {7 t^{2}  - 2 t - 5 \over 12 } {1 \over 2} \delta(p_{0})     
    \langle  \bar{q} q \rangle ^{2}, \label{eq:nucor} \\
    {\rm Im} B^{\rm OPE}(p_{0}) & =  & 
          -  {7 t^{2}  - 2t -5 \over 32 \pi^{2}} p_{0}^{\,2} 
          \theta(p_{0}) \langle \bar{q} q \rangle + 
     {3(t^{2} -1 ) \over 32 \pi^{2}} \theta(p_{0}) \langle \bar{q} g
    \sigma \cdot G q \rangle . \nonumber
\end{eqnarray}
In these expressions we neglect the up quark and down quark masses.  
We allow to determine $t$ so that the current strongly couples to the 
negative-parity states and also require that the contributions of 
higher dimension operators are small in determining $t$.
In Fig.~\ref{fg:t-dep} are plotted the $t$-dependencies of the Borel 
transformed Wilson coefficients of the operators up to dimension 6 
for the nucleon at the Borel mass $M = 1.5$ GeV. Around $t = 1$ and $t 
= -1$, the correction terms of OPE are small compared to the 
identity operator and the coefficients of the higher dimensional 
operators would be small.  In view of the convergence of OPE, such 
$t$ is good for the sum rules.

\begin{figure}[tb]
\epsfxsize=13cm
\epsfbox{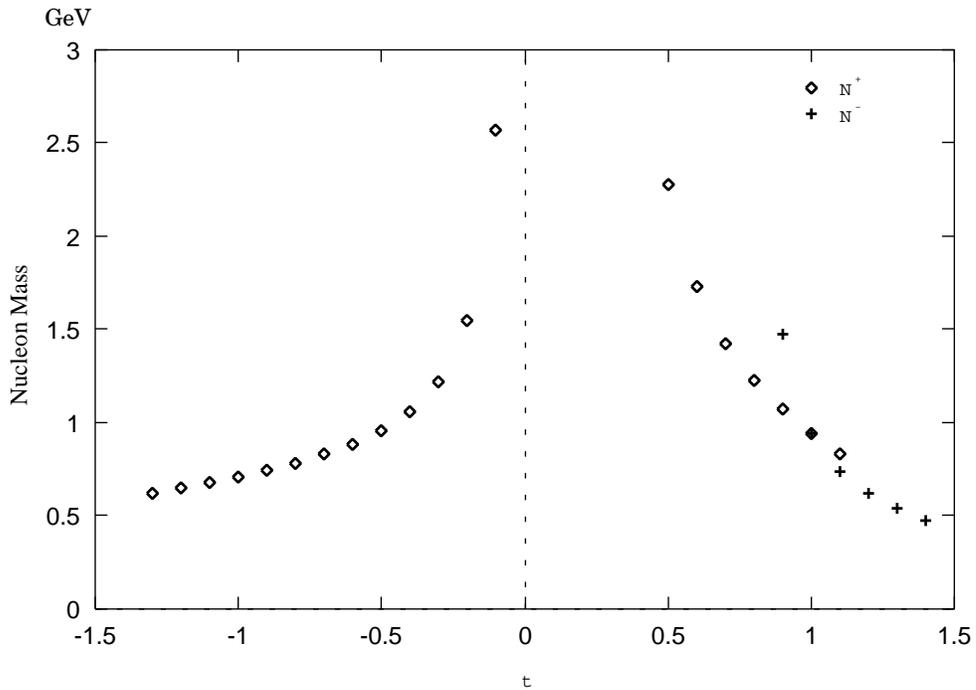}
\caption[]{\small Masses of negative and positive parity nucleons in 
the finite energy sum rule.  The sum rule has no (realistic) solution 
where no dot is plotted.  At $t = 1$ the $N^{+}$ and $N^{-}$ have 
the same mass.  The $N^{+}$ has no solution at $t > 1.1$ and $N^{-}$ 
has no solution at $t<0$.  }
\label{fg:fesr}
\end{figure}

In order to find $t$ such that the current (~\ref{eq:nucur}) couples 
to the negative-parity state, we first apply the finite energy sum 
rule.  It is simple to extract the hadron properties from the finite 
energy sum rule because it contains no additional parameters such as 
the Borel mass.  The results, however, are only qualitatively 
reliable, because they are usually contaminated by higher resonance 
contributions.  Concretely we construct three independent sum rules 
from eqs.(~\ref{eq:sra}) and (~\ref{eq:srb}) choosing three weights 
$W(p_{0})=1$, $p_{0}$ and $(p_{0})^{2}$.  We determine the mass, the 
coupling and the threshold of each parity nucleon by solving the three 
sum rule equations.  The $t$ dependence of the masses of $ N^{+}$ and 
$N^{-}$ are plotted in Fig.~\ref{fg:fesr}.  At $t = 1$, the masses of 
$N^{+}$ and $N^{-}$ are the same because the odd dimensional operators 
($\bar{q}q$, $\bar{q}\sigma\cdot Gq$) do not contribute at $t = 1$ as 
long as we truncate OPE at dimension 6.  There is no solution for 
$N^{-}$ around $t = -1$ although the convergence of OPE would be good.  
This is because the coefficient of the dimension 3 operator 
($\bar{q}q$) is positive and large, so that the current couples weakly 
to the negative-parity states.  If one of the correlation functions of 
$N^{+}$ and $ N^{-}$ is enhanced by the odd dimension operators, the 
other is suppressed, because the contributions from the odd dimension 
operators have different signs for $N^{+}$ and $N^{-}$.  Thus for the 
time being we choose $t = 0.8$.  We shall later study other 
choices of $t$ around 1.

The other input parameters are chosen as
\begin{eqnarray}
    \langle \frac{\alpha_{s}}{ \pi} GG \rangle & = & (0.36 \,{\rm GeV})^{4}  \nonumber \\
  \langle \bar{q} q \rangle & = & (- 0.25 \,{\rm GeV})^{3} \nonumber \\
  \langle (\bar{q} q)^{2} \rangle & = & 1.25 \,\langle \bar{q} q \rangle ^{2} \nonumber \\
  \langle \bar{q} \sigma \cdot G q \rangle & = 
            &(1.0 \, {\rm GeV})^{2} \langle \bar{q} q \rangle \nonumber \\
  m_{u} = & m_{d} & = 0 \nonumber 
\end{eqnarray}
These values are chosen so that the sum rule reproduces the observed 
mass of $N^{+} \simeq$ 940 MeV. We note that the vacuum saturation 
hypothesis that can only be justified in the large $N_{c}$ limit is 
not appropriate for the nucleon sum rule~\cite{nsvvz}.  Indeed, we take 
$\langle(\bar{q}q)^{2}\rangle = 1.25 \langle\bar{q}q\rangle^{2} $, that is, 
25\% enhancement of the four-quark condensate than that with the 
vacuum saturation hypothesis.  
We also note that the ratio of $\langle 
\bar{q} \sigma \cdot G q\rangle$ to $\langle \bar{q}q \rangle$, 
frequently defined as $m_{0}^{\,2} = \langle \bar{q} \sigma 
\cdot G q\rangle / \langle \bar{q}q \rangle $, is consistent with the 
standard value $m_{0}^{\, 2} = 0.5 - 1.0 ({\rm GeV})^{2}$~\cite{rry2}
~\cite{i2}.

We, now, calculate the mass of $N^{-}$ in the Borel sum rule.  We have 
three parameters, the mass, the coupling and the threshold.  Usually, 
one sets up a window in which the QCD sum rules would be effective, 
and then fits the parameters so that they are stabilized with respect 
to the Borel mass in the window.  It is, however, known that the 
results are sensitive to the choice of the window, and therefore have 
significant ambiguity.  Instead we use the following method to fix the 
parameters.  If we choose three arbitrary Borel masses, we can 
determine, in principle, three parameters from the corresponding sum 
rules assuming that those parameters should be independent of the 
Borel mass.  This method should work, if OPE could be summed up to all 
orders.  In practice, however, we can calculate only a few terms of 
OPE and therefore parameters obtained from the sum rule depend on the 
Borel masses.  In order to see sensitivity of the parameters to the 
Borel mass, we select three successive Borel masses each separated by 
$\Delta M = 0.1$ GeV and solve three parameters from the three sum 
rules.  We label the obtained parameters by the center of these Borel 
masses.

\begin{figure}[tb]
\epsfxsize=13cm
\epsfbox{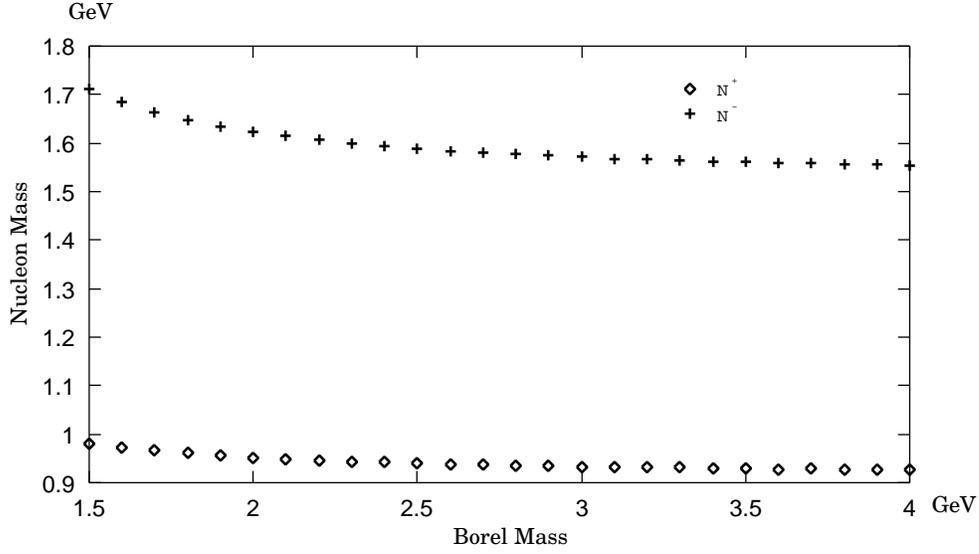}
\caption[]{\small 
     Masses of $N^{+}$ and $N^{-}$ plotted with $\Delta M = 0.1$ 
     GeV.}
    \label{fg:nmass}
\end{figure}

Iterating this procedure, we get the Borel mass dependence of the 
masses of $N^{+}$ and $N^{-}$, plotted in Fig.~\ref{fg:nmass}.  Both 
masses are almost independent of the Borel mass.  Our sum rule gains 
extra stability due to the integral measure.  The integral measure 
$dp_{0}^{2}$ in the standard sum rule is $2p_{0}dp_{0}$ in ours, which 
adds a $p_{0}$ enhancement of the continuum term, and makes the pole 
contribution to the sum rule weaker.  By fixing the QCD parameters so 
as to give the $N^{+}$ mass $\sim$ 940 MeV we predict the $N^{-}$ mass 
about 1550MeV. The observed $N^{-}$ mass is 1535MeV with the width of 
150MeV $\pm$ 15MeV~\cite{pd} and our prediction agrees very well.  Note 
that the $N^{+}-N^{-}$ mass difference is caused by the terms of the 
odd-dimension condensates ($\langle\bar{q}q\rangle$, 
$\langle\bar{q}\sigma\cdot Gq\rangle$).  Thus one might say that QCD 
chooses $N^{+}$ as the ground state by setting $\langle \bar{q}q 
\rangle < 0$.

We check the dependence of the results on $\Delta M$ and $t$.  We 
calculate the masses of $N^{+}$ and $N^{-}$ with $\Delta M = $ 0.05, 
0.2 and 0.5 GeV in the same way.  Although at around $M = 1.5$ GeV the 
mass obtained from each sum rule is different from the others by a few 
per cent, the masses are stabilized above $M = 2.5$ GeV and each sum 
rule gives the same result.
In order to study the $t$ dependence, we calculate the nucleon masses with 
other choices of $t =$ 0.9, 1.05, and 1.1.  For each $t$ we adjust 
the QCD parameters so that the $N^{+}$ mass is reproduced.  For 
$t = 0.9$ the $N^{-}$ mass is about 1.4 GeV, and the mass 
difference of $N^{+}$ and $N^{-}$ becomes smaller for $t = 0.8$.  This 
is because the contribution from the odd dimensional operators is 
larger for larger $t$.  We are interested in the case $t > 1$ because 
in this region the dimension 3, 5 and 6 operators have the opposite 
sign to those for $t<1$.  Then the mass of $N^{-}$ could be smaller 
than that of $N^{+}$.  But we find that for $t =$ 1.05 and 1.1 the 
$N^{-}$ sum rule has no solution when the QCD parameters are adjusted 
so as to give the mass of $N^{+}$.  The current with $t > 1$ seems not 
to couple with the negative-parity nucleon.

\section{Roles of the Chiral Symmetry Breaking}
As we see in Sec.~\ref{sc:sep}, the difference of the masses of the 
positive-parity baryon and the negative-parity baryon is caused by 
the chirally odd term $B$ in eq.(~\ref{eq:imphe}).  In this section, 
we study how the terms breaking the chiral symmetry determines the 
parity splitting.

In the correlation function of the nucleon eq.(~\ref{eq:nucor}) the 
chiral symmetry is broken by the vacuum expectation values of the 
operators $\bar{q}q$,$\bar{q} \sigma \cdot G q$ and a part of 
$\bar{q}q\bar{q}q$.  The first two are split into two chiral 
terms,
\begin{eqnarray}
  \langle\bar{q}q\rangle & = & \langle\bar{q_{L}}q_{R}\rangle +
   \langle\bar{q_{R}}q_{L}\rangle,\\
  \langle\bar{q}\sigma\cdot Gq \rangle & = & 
  \langle\bar{q_{L}}\sigma\cdot Gq_{R}\rangle + 
  \langle\bar{q_{R}}\sigma\cdot Gq_{L}\rangle,
\end{eqnarray}
and each term breaks the chiral symmetry.  The vacuum expectation 
value of the four-quark operator $\bar{q}q\bar{q}q$ in the nucleon can 
be written as a sum of three terms with different chiral properties, 
\begin{equation}
 (7t^{2}-2t-5) \langle\bar{q}q\bar{q}q\rangle = (t^{2}-2t+1)\, 
 4\langle\bar{q_{L}}q_{R}\rangle \langle\bar{q_{R}}q_{L}\rangle 
 -6(1-t^{2})\,2(\langle\bar{q_{L}}q_{R}\rangle^{2} + 
 \langle\bar{q_{R}}q_{L}\rangle^{2}).  \label{eq:dim6ch} 
\end{equation}
Note that we use the vacuum saturation hypothesis~\cite{svz} and this formula is 
only for the nucleon sum rule.  The first term is the chiral symmetric 
term because the term breaks the chiral symmetry twice and the net 
chirality is preserved.  In the second term the chirality is broken.  
With our choice of $t=0.8$, eq.(~\ref{eq:dim6ch}) breaks the chiral 
symmetry strongly since the chiral noninvariant term is dominant.  If 
we choose $t=-1$, the term breaking chiral symmetry vanishes. So the 
$\langle\bar{q}q\bar{q}q\rangle$ term for the Ioffe's current ($t=-1$) 
is invariant under the chiral symmetry.

In order to see the effect of the chiral symmetry breaking, we vary 
$\langle \bar{q} q \rangle $ and study its effects.  $\langle \bar{q} 
\sigma \cdot G q \rangle$ is assumed to be proportional to $\langle 
\bar{q} q \rangle$ and therefore is varied together with $\langle 
\bar{q} q \rangle$.  As eq.(~\ref{eq:dim6ch}) $\langle 
\bar{q}q\bar{q}q\rangle$ is reduced to the square of $\langle \bar{q} 
q \rangle$ and we vary only the terms breaking the chiral symmetry.  
We define the ratio $R$ of $\langle \bar{q} q \rangle $ to its 
standard value $\langle \bar{q} q \rangle _{0}$, $R = \langle \bar{q} 
q \rangle / \langle \bar{q} q \rangle_{0}$, and choose seven values of 
$R = 0.3,0.5,0.8,1.0,1.2,1.5,2.0$.

\begin{figure}[tb]
\epsfxsize=13cm
\epsfbox{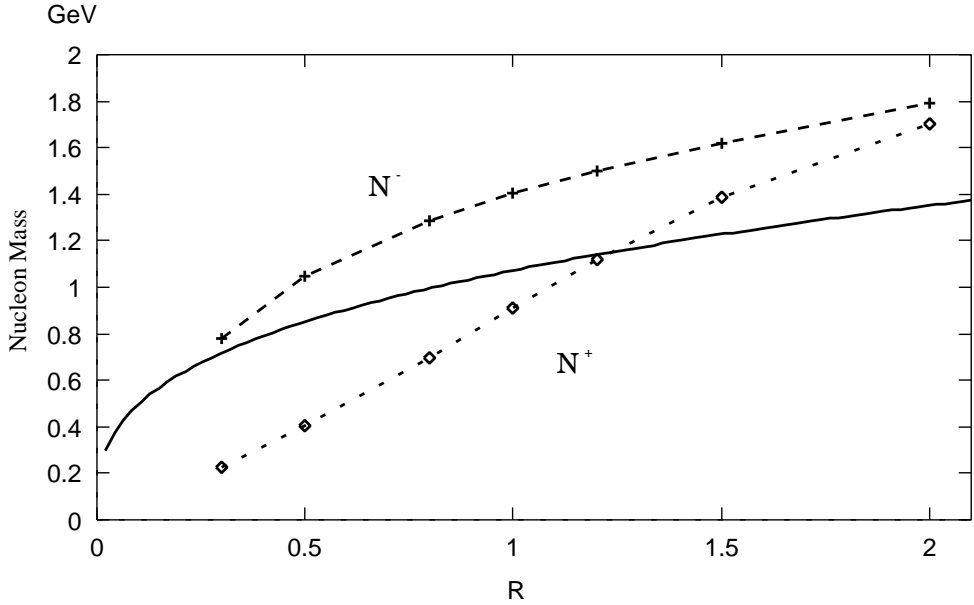}
	 \caption[]{\small Masses of $N^{+}$ and $N^{-}$ at $M=2.5$ GeV 
	 for various values of the quark condensate.  $R$ is the ratio of 
	 $\langle \bar{q} q \rangle $ to its standard value $\langle 
	 \bar{q} q \rangle _{0}$.  The solid line is the Ioffe's formula 
	 (~\ref{eq:ioffo}).  }
    \label{fg:nvarq}
\end{figure}

In Fig.~\ref{fg:nvarq} are plotted the masses of nucleons with 
positive and negative parity.  One sees that both the masses of 
$N^{+}$ and $N^{-}$ go towards zero when the ratio $R$ goes to zero.  
Although we cannot confirm that the $N^{+}$ and $N^{-}$ masses go to 
zero at $R \rightarrow 0$, they should be degenerate because the 
chiral symmetry breaking term $B$ vanishes.  Both of the $N^{+}$ and 
$N^{-}$ masses grow for large $R$ and become degenerate, but $R$ 
dependent behaviors are different.  It should be noted that the 
behavior of the $N^{+}$ mass is different from the Ioffe's 
formula~\cite{i}
\begin{equation}
m^{+} = [-2 (2\pi)^{2} \langle \bar{q}q \rangle]^{1/3} \label{eq:ioffo}
\end{equation}

The reason why the $N^{+}$ and $N^{-}$ masses at large $R$ become 
degenerate is as follows.  Both masses grow with increasing $R$ and 
the thresholds are simultaneously enhanced.  The enhancement of the 
threshold makes the bare loop term dominant to the others.  The 
correlation functions of $N^{+}$ and $N^{-}$ become similar since the 
bare loop term contributes to the correlation functions with the same 
sign, and therefore the masses tend to be degenerate.  Behaviors of 
the $N^{+}$ and $N^{-}$ masses vs.\ $R$ can be explained by realizing 
that the $\langle \bar{q} q\rangle^{2}$ term (dimension 6) enhances 
the baryon masses. This is related to the sign of $\langle 
\bar{q}q\rangle^{2}$ term, that is negative for both the $N^{+}$ and 
$N^{-}$ sum rules. Thus the $\langle \bar{q} q\rangle^{2}$ term is 
suppressed when $R$ grows and the sum rule tends to increase 
$s_{0}^{\pm}$ to compensate its effect. As a result, the baryon mass 
increases. On the other hand, the $B$ term ($\langle \bar{q}q 
\rangle + \langle \bar{q} \sigma \cdot G q \rangle$) contributes with 
different signs to $N^{+}$ and $N^{-}$. For $N^{+}$, it tends to 
suppress the mass increase around $0.3\leq R \leq 1.0$, while $N^{-}$ 
mass is enhanced there. Thus the $N^{+}$ mass grows slowly in 
comparison with the $N^{-}$ mass.

\section{Summary}

The interpolating current for the octet baryons couples also to the 
negative-parity baryons.  We separate the contribution of the 
negative-parity baryon from the sum rule for the positive-parity 
baryon.  Using the particular correlation function (~\ref{eq:ourco}), 
we construct the sum rule for the negative-parity baryons in the 
$p_{0}$ complex plane.

We apply the formulation to the masses of $N^{-}$.  We obtain the 
Borel mass stability in a wide region for both the $N^{+}$ and $N^{-}$ 
masses.  We find that the negative $\langle \bar{q}q \rangle$ 
condensate gives a heavier $N^{-}$ mass than $N^{+}$ mass.  We find 
that the current~\ref{eq:nucor} with $t \simeq 0.8$ couples 
strongly to the negative-parity state.

In order to see the roles of the chiral breaking to the parity 
splitting, we study the behavior of the $N^{+}$ and $N^{-}$ masses by 
varying the quark condensate.  The smaller the quark condensate 
we use, the smaller the masses are, and the $N^{+}$ and $N^{-}$ masses 
are degenerate when the quark condensate vanishes.  At a larger quark 
condensate the masses are also degenerate, since the dimension 6 
operator plays the dominant role there.

\pagebreak[4]

\begin{thebibliography}{9}
    \bibitem{svz} M.A. Shifman, A.I. Vainshtein and V.I. Zakharov, 
    Nucl. Phys. B147 (1979) 385, 448.
    \bibitem{rry}  L.J. Reinders, H. Rubinstein and S. Yazaki, 
	   Phys. Rep. \underline{127} No.1 (1985) 1.
    \bibitem{i} B.L. Ioffe, Nucl. Phys. B188 (1981) 317, (E) B191 
    (1981) 591.
    \bibitem{i2} V.M. Belyaev and B.L. Ioffe, Sov. Phys. JETP 56 (1982) 493.
    \bibitem{ept} D. Espriu, P. Pascual and R. Tarrach, Nucl. Phys. 
B214 (1983) 285.
    \bibitem{cdks} Y. Chung, H.G. Dosch, M. Kremer and D. Schall, Nucl. 
Phys. B197 (1982) 55.
    \bibitem{nsvvz} V.A. Novikov, M.A. Shifman, A.I. Vainshtein, M.B. Voloshin
and V.I. Zakharov, Nucl. Phys. B237 (1984) 525.
    \bibitem{rry2} L.J. Reinders, H.R. Rubinstein and S. Yazaki, Phys. 
Lett. 120B (1983) 209.
    \bibitem{pd} Particle Data, Phys. Rev. D50 (1994) 1691
\end{thebibliography}
\end{document}